\documentclass{article}
\usepackage{Arxiv_Styl}
\usepackage{amsmath}
\usepackage[utf8]{inputenc} 
\usepackage[T1]{fontenc}    
\usepackage{hyperref}       
\usepackage{url}            
\usepackage{booktabs}       
\usepackage{amsfonts}       
\usepackage{nicefrac}       
\usepackage{microtype}      
\usepackage{lipsum}
\usepackage{fancyhdr}       
\usepackage{graphicx}       
\usepackage{xcolor}
\graphicspath{{media/}}     
\usepackage{multicol}
\usepackage{longtable}
\usepackage{chemformula}
\usepackage{array}
\usepackage[backend=biber]{biblatex}
\immediate\write18{cp `kpsewhich biblatex-dm.cfg` ./biblatex-dm.cfg}

\title{Who Waits the Longest? Unequal Paths to Legal
Stability among Migrants and Refugees}

\author{
  Ola Ali \\
  Complexity Science Hub \\
  \texttt{ola@csh.ac.at} \\
  \And
  Elma Dervic \\
  Complexity Science Hub \\
  Supply Chain Intelligence Institute Austria \\
  \texttt{dervic@csh.ac.at} \\
  \And
  Guillermo Prieto-Viertel \\
  Complexity Science Hub \\
  \texttt{prieto-viertel@csh.ac.at} \\
  \And
  Carsten Källner \\
  Complexity Science Hub \\
  \texttt{kaellner@csh.ac.at} \\
  \And
  Rainer Stütz \\
  Complexity Science Hub \\
  \texttt{stuetz@csh.ac.at} \\
  \And
  Andrea Vismara \\
  Complexity Science Hub \\
  \texttt{vismara@csh.ac.at} \\
  \And
  Rafael Prieto-Curiel \\
  Complexity Science Hub \\
  \texttt{prieto-curiel@csh.ac.at} \\
}

\begin{document}

\maketitle

\maketitle

\begin{abstract} 
Legal systems shape not only the recognition of migrants and refugees but also the pace and stability of their integration. Refugees often shift between multiple legal classifications, a process we refer to as the ``legal journey''. This journey is frequently prolonged and uncertain. Using a network-based approach, we analyze legal transitions for over 350,000 migrants in Austria (2022–2024). EU nationals such as Germans reach stable residence in under four months; refugees from conflict-affected regions face far longer and more uncertain journeys, ranging from two months for Ukrainians to nine months for Syrians and 20 months for Afghans. Refugee women, especially from Syria and Afghanistan, are more likely to gain protection. Afghan women reach stability in 14 months on average, less than half the 30 months Afghan men wait. We also find that those who cross the border without going through official border controls face higher exit rates and lower chances of securing a stable status. We show that legal integration is not a uniform process, but one structured by institutional design, entry mode, and unequal timelines.

\end{abstract}

\section{Introduction}

{
As of 2024, around 40 million individuals worldwide are considered refugees, and approximately 8 million are currently seeking asylum \cite{unhcr2024stats}. In Europe alone, over 2.8 million new asylum applications were filed between 2022 and 2024, with more than 1 million in 2023 alone, the highest number since the 2015-2016 refugee crisis. Austria, a key destination and transit country, received over 187 thousand asylum applications during this period~\cite{eurostatA, AIDA_Austria_Statistics}. These figures reflect more than just policy trends or administrative burdens. They represent millions of lives suspended in legal uncertainty, waiting for decisions that profoundly shape their futures, where forced stillness becomes the defining condition of their existence~\cite{Della_Puppa}. 
}

{
Legal status is a defining element in the lives of refugees and migrants. It determines where they can live \cite{stability_Refugee}, whether they can work \cite{ebner2023}, and contribute to their host communities \cite{hainmueller2016lives, bertrand2019refugees}. More than a bureaucratic label, legal status sets the conditions under which individuals can integrate \cite{bakker2014importance}. Yet for many, this process is neither quick nor straightforward. Obtaining a stable legal status is a lengthy and complex process, largely dictated by host country policies, which can take several months to years to complete. 
}

{
Despite its relevance, legal status is rarely treated as a dynamic process in quantitative migration research. However, qualitative and anthropological traditions have long examined the legal dimension as a lived, evolving experience~\cite{eule2018migrants}, with particular attention to how waiting and temporal uncertainty shape the lives of migrants and refugees~\cite{storato2023timescapes}. Existing studies on integration often focus on end outcomes, such as employment \cite{hainmueller2016lives, aaslund2024limbo}, healthcare \cite{hvidtfeldt2020prolonged, phillimore2021violence}, education \cite{gladwell2018education, sirin2015educational}, and social benefits \cite{heredia2022missing}, without accounting for the transitional legal phases that precede them~\cite{munz2007migration,dustmann2011migration,dervic2024healthcare}. Much of the existing research tends to focus on aggregated statistics \cite{abel2014quantifying,abel2010estimation}, and in the case of refugees, acceptance or rejection rates, offering limited insight into the complex, evolving nature of legal status transitions~\cite{AIDA_Austria_Statistics,biffl2024migration}. This overlooks a critical period of vulnerability, especially for refugees. Delays or uncertainties in legal processing can exacerbate housing instability \cite{stability_Refugee}, hinder access to employment \cite{organisation2016making,martin2016refugees}, increase risks of psychiatric disorders \cite{hvidtfeldt2020prolonged}, and undermine social inclusion. Additionally, research has shown that long waiting periods in the asylum process contribute to feelings of powerlessness and immobility, particularly among vulnerable groups like children \cite{Vianelli04042022, wolter2023waiting}. These delays also disrupt family reunification efforts \cite{UNHCR_Family_Reunification}, further complicating integration and affecting the overall well-being of refugees \cite{ hvidtfeldt2022waiting, kohlenberger2019barriers}. Gender further shapes these legal trajectories, recent findings point to a  ``female advantage'' in asylum outcomes, where females have a slightly higher acceptance rate \cite{ortensi2024female}. Beyond these outcomes, prolonged waiting functions as a mechanism of temporal governance that disciplines migrants and constrains their ability to plan, work, and build social ties~\cite{Della_Puppa,sano2024claiming}. 

}

{
Austria provides a particularly relevant case for examining legal status transitions, where refugees are around 17\% of the total migrant population \cite{stability_Refugee}. While the country no longer follows a quota system for refugee admissions, its approach to migration and integration has evolved over the years \cite{Borkert2015} and continued to provide significant support to asylum seekers, assisting 18,273 refugees in 2018 with a financial commitment of over EUR~54~million~\cite{OECD_ODA_Refugee_Costs_Austria}. Despite these efforts, the legal pathways available to migrants and refugees remain slow and uneven. 
A fundamental structural distinction shapes these legal pathways one that is embedded in law before any individual application is filed. EU nationals enjoy freedom of movement under Directive 2004/38/EC and are not required to obtain authorisation to reside~\cite{Directive2004_38}, whereas third-country nationals must apply for and be granted residence rights under national law~\cite{NonEU_first_2026}. For asylum seekers, the applicable framework is the Geneva Convention and its EU-level implementation~\cite{geneva-refugee-convention,LukasAIDA2024}; additional instruments, such as the Temporary Protection Directive, have been activated for Ukrainians following the 2022 Russian invasion~\cite{TempAIDA2024}. These structural differences mean that legal complexity is not evenly distributed across migrant and refugee populations. Yet how these structural differences translate into individual legal trajectories, and how their duration and outcomes vary by country of origin, gender, and entry status, remains poorly understood. This lack of understanding hinders efforts to improve integration strategies through legal support.

}

{
This study investigated how legal status transitions among migrants and refugees in Austria vary by country of origin and what the implications are for their long-term legal integration and stability. We apply network analysis to administrative records of all migrants and refugees in Austria who registered their residence between 2022 and 2024 in the central register of residents (Zentrales Melderegister/ZMR), examining transitions between legal statuses over a two-year period. In total, our data covers 358,327 
individuals, 70\% of whom changed legal status at least once during the observed period. By constructing a network of legal pathways (Fig.~\ref{fig:Network}), we map the 
transitions between legal statuses for each country of origin. We assume a Markovian model to capture the time to stability and predict future trajectories. These trajectories are shaped by factors such as the migrant’s country of origin, gender, and the host country’s legal frameworks.
}

{ 
We find that refugees from conflict-affected areas, such as Afghanistan and Syria, face fragmented and prolonged legal journeys with significantly more transitions, while migrants from countries like Germany or Italy follow straightforward legal trajectories, achieving stability quickly. Gender also plays a role where women, particularly from Syria and Afghanistan, achieve stable legal statuses with higher probability, while men face prolonged instability. These findings highlight that legal transitions are not merely case-by-case occurrences but are shaped by systemic factors. 
}
\section{Results}

{
A typical legal journey of migrants upon entering the country involves applying for a residence permit and receiving it within a few weeks. Once the purpose of the permit ends, such as study or employment, the individual may either leave or transition to another status. Refugees, on the other hand, undergo a much longer and more complex process, starting with applying for asylum, which may be subject to rejection and appeal at various stages, before ultimately obtaining asylum and settling in Austria (Fig.~\ref{fig:Walk_legal}).

{
\begin{figure}[ht]
    \centering
    \includegraphics[width=0.8\linewidth]{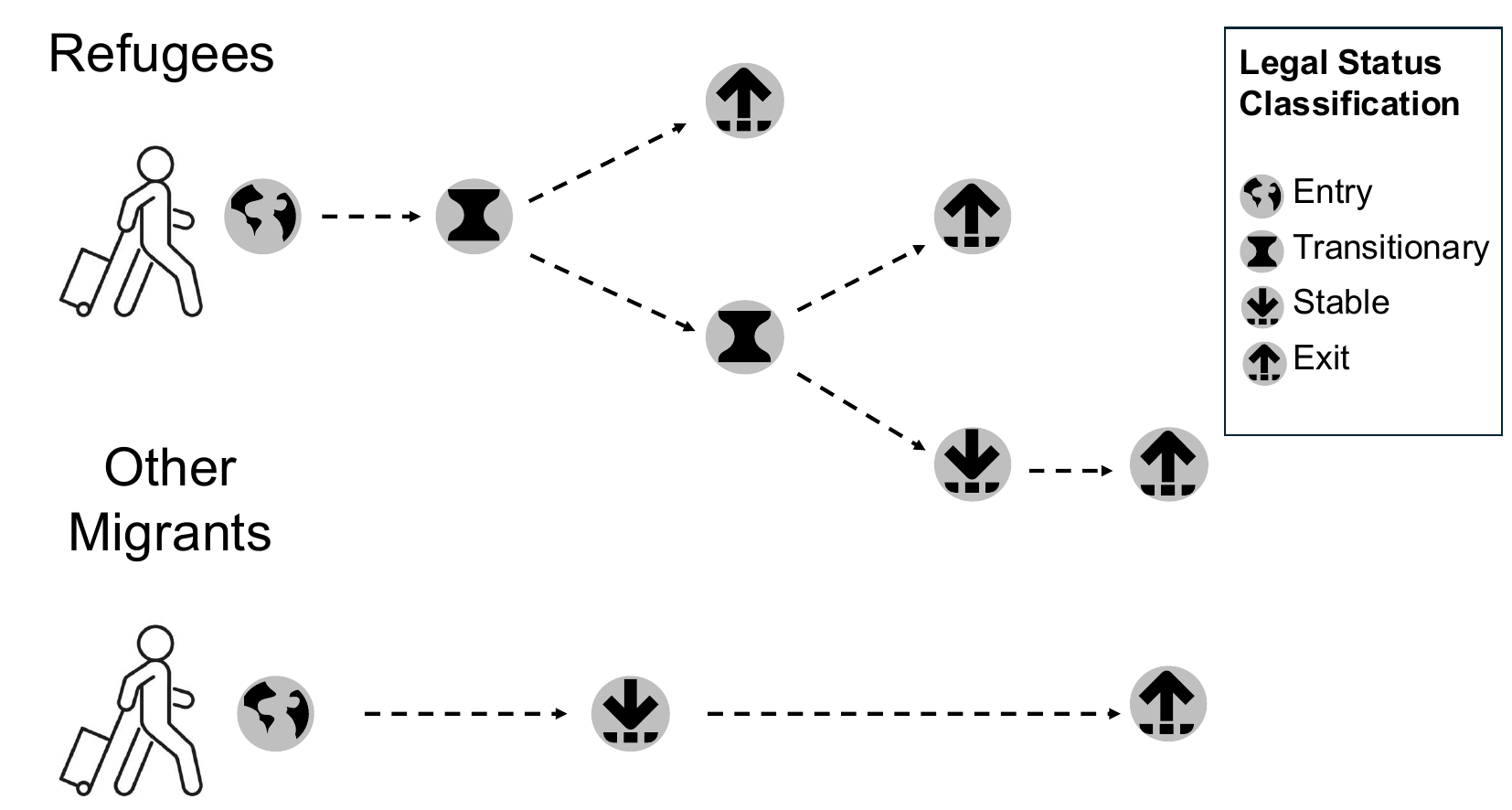}
    \caption{\textbf{A Walk Through The Legal System}. This illustration shows a refugee's average legal transition journey (Top) from a foreigner, applying for asylum, being approved as an asylum seeker, to being entitled to asylum, to exit. Compared with a migrant (Bottom), going from the status of a foreigner to having a residence permit, to exit.}
    \label{fig:Walk_legal}
\end{figure}
}
}

{ 
Over the two-year observation period, we analyzed how migrants move from one legal status to another. The Austrian legal system defines 11 official legal statuses, to which we add an ``exit'' status to account for individuals who have legally left the country. The dataset includes over 350,000 migrants and refugees, and more than 140 million daily legal status records, capturing transitions across entry-level, transitionary, stable, and exit-related statuses. We define a stable legal status as one that grants long-term residence and access to rights, including work, healthcare, and protection from removal. 

Each country of origin exhibits a unique legal transition pattern, reflecting the behaviors and legal pathways of its migrants in Austria (Fig.~\ref{fig:Network}). We refer to them as ``signatures''. Migrants from countries like Germany, Italy, or Croatia, generally display the simplest signature with the most transitions occurring between entry, obtaining a stable status ``residence permit'', and eventual exit. A similar pattern is observed in migrants from countries like Kosovo and China. 

}
{
Countries with ongoing conflicts have complicated dynamics, where the signatures are more complicated, like Syria and Afghanistan. For these countries, the most common outcomes are stable statuses such as ``entitlement to asylum'' or ``subsidiary protection''. By contrast, although the path to stability is less complicated, Ukrainian migrants more often end up with the status of ``displaced person'', which under the Displaced Persons Act applies to those who lived in Ukraine before 24 February 2022~\cite{AIDA-Temp-AT2022}. 
For Russia, the number of migrants with the status ``foreigner with obligation to leave the country'' is considerable compared to the rest.
\begin{figure}[!ht]
  \centering
  \includegraphics[width=0.8\textwidth]{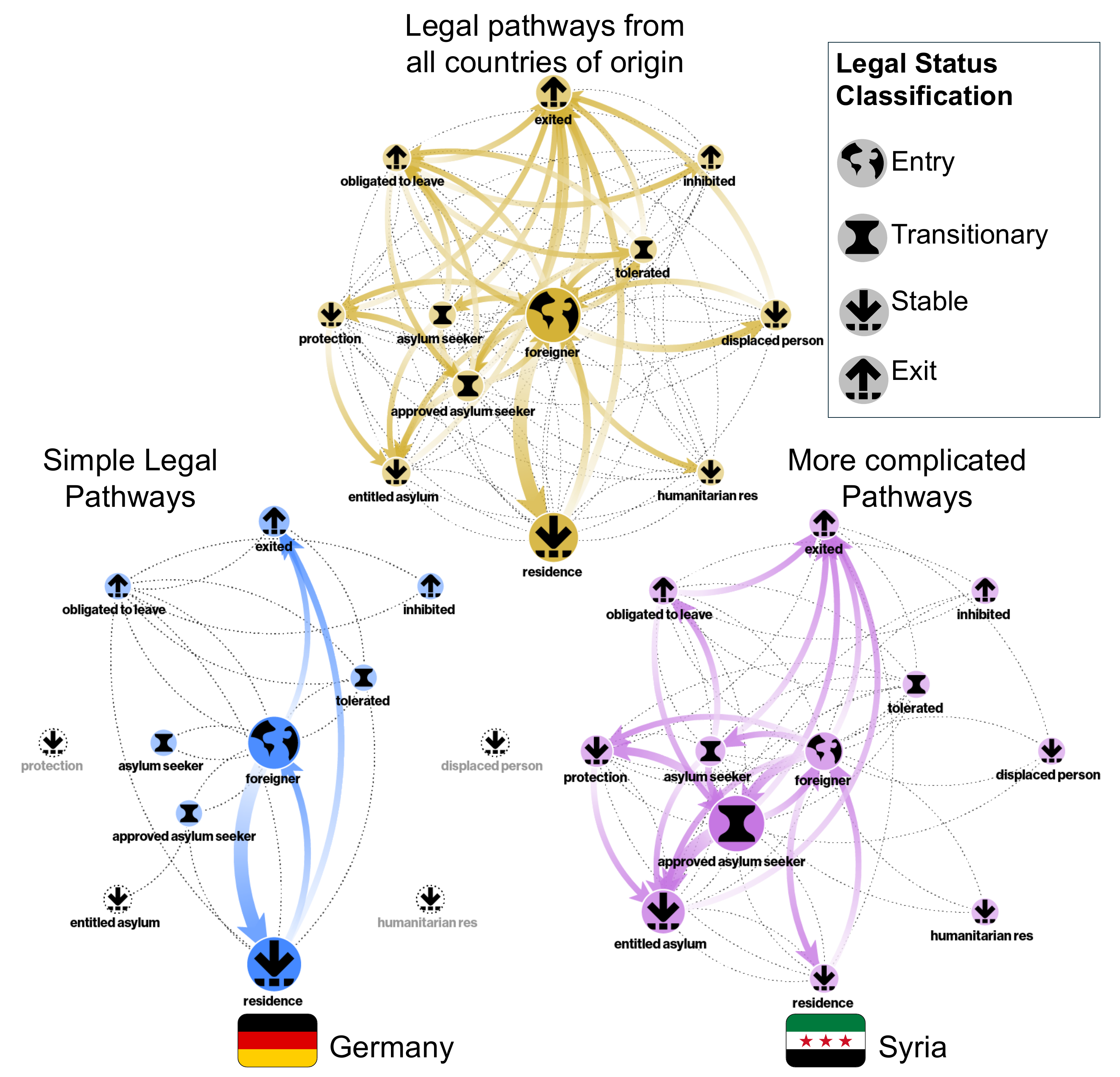}
  \caption{\textbf{Legal Status Network of Migrants in Austria}. Nodes correspond to legal statuses, their size corresponds to the total incoming and outgoing migrants, and dashed nodes have no migrants. Edge width represents the daily average number of migrants transitioning between statuses, and dashed edges indicate fewer than 100 migrants. This figure illustrates the unique signatures of legal transitions for the two largest migrant and refugee populations in Austria. The network for all countries is available at: \url{https://vis.csh.ac.at/migration-legal-status/}.}
  \label{fig:Network}
\end{figure}
}

{
While the signatures reveal broad patterns in legal trajectories across countries, they do not capture how quickly individuals reach legal stability. Time plays a critical role in shaping migrants' access to rights, services, and long-term integration prospects. To assess the temporal dimension, we model legal transitions over two key time frames: the first year, when migrants are typically expected to secure their legal footing, and a decade, to capture longer-term stability patterns. Modeling is necessary because we often observe only direct switches between statuses, without full legal trajectories. It also enables us to infer likely future pathways and compare legal outcomes across groups, even when the data are incomplete (more information in SI part 1).
}

\subsection{Reaching Stability: First Year Post Arrival}

{
Within the first year of arrival, migrants are generally expected to achieve a stable legal status, which is often necessary for accessing long-term residence and labor market opportunities. While legal requirements differ across national contexts, such as whether formal permits are needed to work, the most meaningful variation in legal outcomes occurs at the country level, not along broad regional lines, such as whether a country is part of the European Union (EU) or not (more information in SI part 3). The proportion of migrants who attain stability and the speed at which they do so vary considerably by country of origin. To explore these differences, we model a population of 10,000 migrants from each country and illustrate the distribution of legal statuses (Fig.~\ref{fig:OneYear}). We focus on the three largest migrant groups (Germany, Turkey, and Serbia) and the three largest refugee groups (Syria, Ukraine, and Afghanistan) in Austria, based on the size of their populations, to capture the most relevant differences in legal stability~\cite{prieto2023diaspora,stability_Refugee}. 
}

{
In the case of Germany, approximately 80\% of migrants obtain a residence permit, 15\% leave the country, and the remaining 5\% retain a foreigner status. For Turkey, fewer than half (49\%) achieve residence, around 35\% leave, and a small portion (approximately 7\%) have their asylum applications accepted, though notably, none ultimately receive formal asylum status. The rest remain as foreigners. Serbia presents another striking case, where the share of those leaving (39\%) is nearly equal to those obtaining residence (40\%), with the remainder staying in a ``foreigner'' status.
}

{
For Syria, over half of the migrants attain a stable refugee status. Specifically, about 56\% are granted ``entitled to asylum'' status, while 13\% receive ``subsidiary protection''. Around 14\% remain in the asylum process, and the rest either obtain residence permits or leave the country. This stands in contrast to Afghanistan, where the largest share (30\%) obtains a residence permit. Approximately 27\% receive refugee status, split between 16\% ``entitled to asylum'' and 11\% ``subsidiary protection''. Ukraine follows a distinct pathway; the majority of Ukrainian migrants (68\%) are classified as ``displaced persons,'' while nearly all the remaining (28\%) exit the country.

\begin{figure}[ht!]
  \centering
  \includegraphics[width=\textwidth]{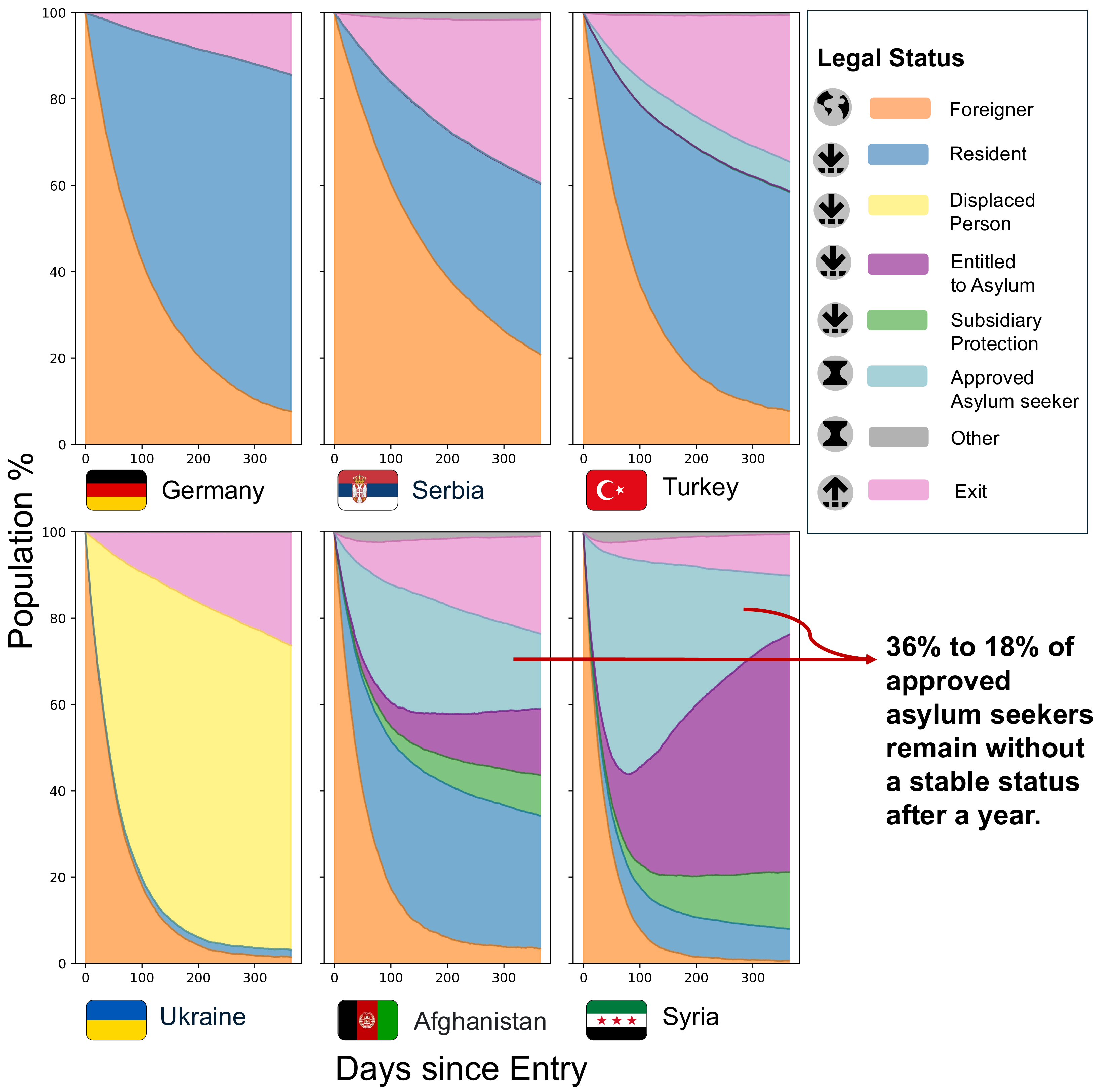}
  \caption{\textbf{Population Legal Distribution within the First Year of Entry.} Modeled share of 10,000 migrants from various countries of origin who enter as foreigners and transition through different legal statuses. The probability of attaining stability within the first year is estimated based on observed transition patterns. An interactive version of this plot of all countries of origin is available at: \url{https://vis.csh.ac.at/migration-legal-status/}.}
  \label{fig:OneYear}
\end{figure}
}

\subsection{Alternative Entry: Entering as an Asylum Seeker Vs Foreigner}
{
The entry status of asylum seekers, whether they first enter as foreigners or cross the border without going through official border controls to apply for asylum, plays a critical role in determining the likelihood of asylum acceptance or rejection. In our dataset, the majority of refugees enter the country as foreigners; however, a small subset enters as asylum seekers. This latter group faces a significantly higher probability of exiting the country within the first year. For instance, for Syria, asylum seekers have a 45.7\% chance of exiting, compared to only 9.6\% for those entering as foreigners. For Afghanistan, the probability is 54\% vs. 22.5\%, and in Ukraine, it is 60\% vs. 26.3\% (Fig.~\ref{fig:Alternative_Entry}). This demonstrates that exit rates are two to four times higher, depending on both the mode of entry and the country of origin.

The higher exit probability corresponds to a lower likelihood of obtaining stable asylum status. For Ukraine, the probability of gaining asylum drops from 70\% (for those entering as foreigners) to 31\% (for those entering as asylum seekers). For Syria, it decreases from 68\% to 42\%, and for Afghanistan, it falls from 25\% to 23.3\%.

After one year of entry, the percentage of approved asylum seekers still awaiting a decision (whether positive or negative) shows minimal variation based on entry status. For example, for Afghanistan, the share of approved asylum seekers after one year is approximately 17\%, compared to 17.5\% for those who entered as foreigners. For Syria, the share is 10\% for those entering as asylum seekers vs. 13.6\% for those entering as foreigners. The most notable difference occurs for Ukraine, where the share of individuals remaining as approved asylum seekers is almost nonexistent for those who entered as foreigners (0.02\%), while it stands at 4\% for asylum seekers.

\begin{figure}[!ht]
  \centering
  \includegraphics[width=\textwidth]{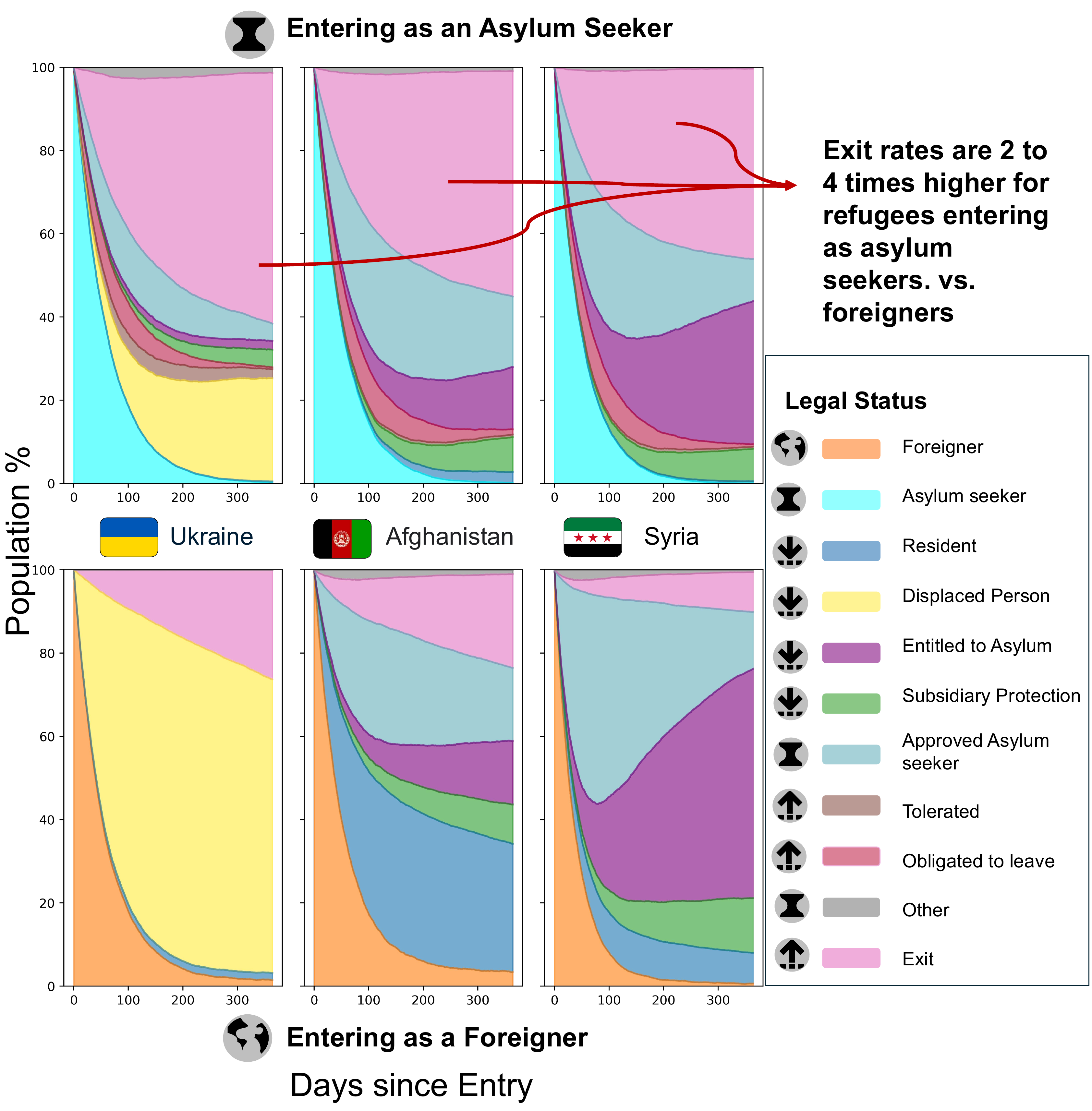}
  \caption{\textbf{Entering as an Asylum Seeker vs Foreigner.} Modeled share of 10,000 migrants from conflict-affected countries of origin who enter as asylum seekers (Top) vs as foreigners (Bottom) and transition through different legal statuses. The probability of attaining stability within the first year is estimated based on observed transition patterns.}
  \label{fig:Alternative_Entry}
\end{figure}
}
\subsection{Long Term Outlook}

{
Migrants, especially those from conflict-affected countries, require more time to navigate the system and achieve a stable legal status. To account for these longer trajectories, we extend our simulation to cover a decade, allowing us to estimate the average time needed to reach legal stability for each group. We find that the average time required for a migrant to attain a stable status (i.e., ``residence permit'') within this period varies notably by country of origin. In the case of Germany, migrants reach stability the fastest and the most, with 3.7 months average and around 94\% settle, followed by those from Turkey with 3.9 months and around 71\%, and Serbia around seven months and around 63\% (Fig.~\ref{fig:stabilityTime}). 
}
{
In contrast, for countries affected by conflict, where obtaining a stable asylum status is essential for accessing housing, employment, and other benefits, we simulate the same population size, tracking individuals from entry as foreigners to reaching one of the recognized asylum statuses (``displaced person'', ``entitled to asylum'', ``subsidiary protection'', or a ``humanitarian residence permit''). Here, Ukrainian refugees have the shortest average time to reach stability, with just around two months to attain the ``displaced person'' status. This is followed by refugees from Syria at almost nine months with 89\%, while refugees from Afghanistan face the longest wait and settle the least, averaging around 20 months with 56\% of the population.

\begin{figure}[ht!]
    \centering
    \includegraphics[width=0.8\linewidth]{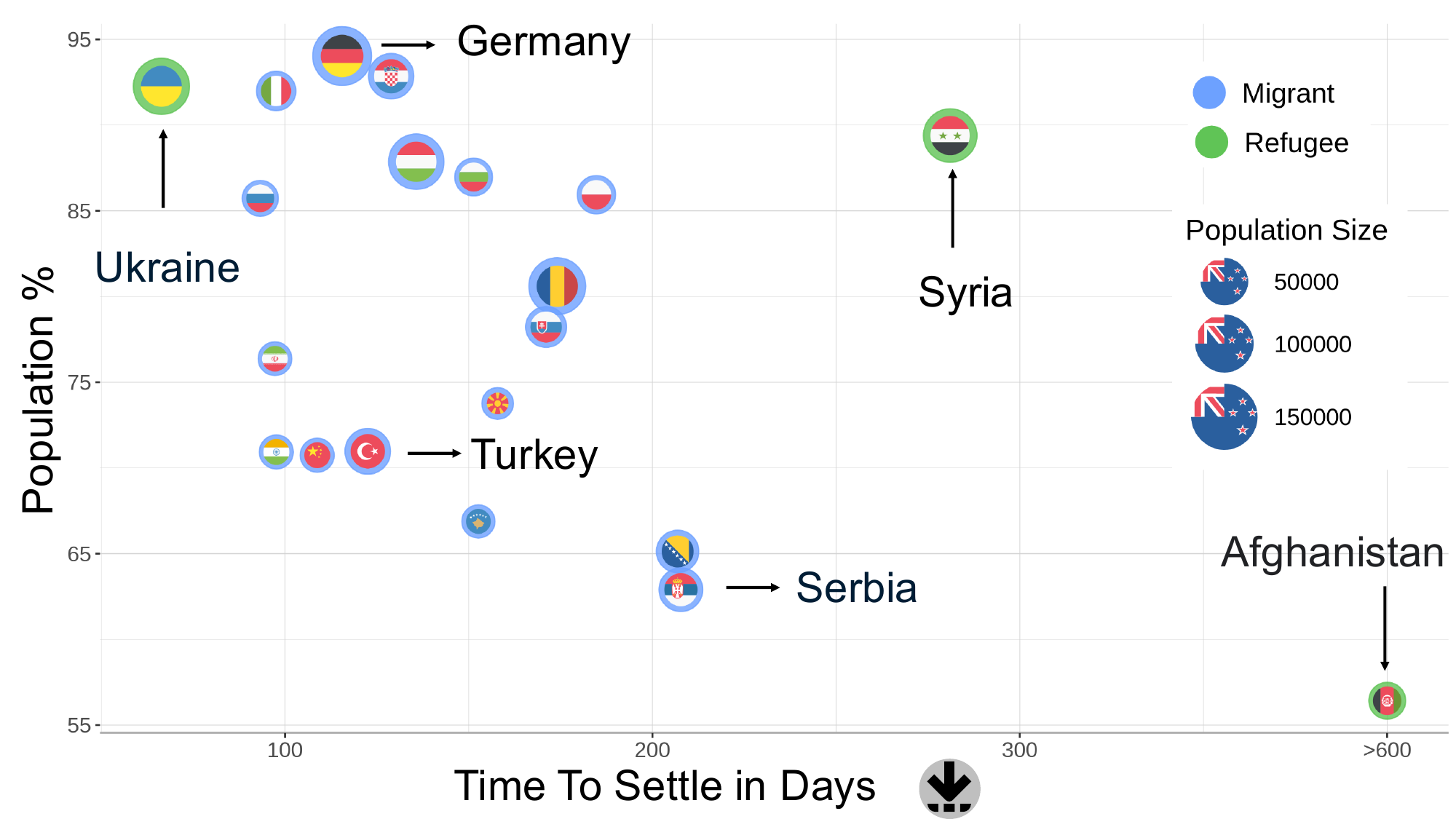}
    \caption{\textbf{Settling Time within a Decade of Entry.} The expected time in days to obtain a stable status for migrants (which is residence permit) and refugees (one of the four refugee statuses, more on the stable statuses in Table~\ref{tab:legal_classes}) on the horizontal line while the percentage of the population that reaches the stability within the decade are on the vertical line. }
    \label{fig:stabilityTime}
\end{figure}
}

{
We investigate how gender affects the observed patterns in legal outcomes (more information on gender in SI part 2). For some countries, there is some ``female advantage'', meaning that females have a slightly higher probability of gaining asylum. In the case of Afghanistan, for example, women are more likely to gain asylum (20\% vs 14\% ) and reach it faster (14 vs. 30 months). Additionally, the average duration of the legal journey, from entry to either exit or stability to exit, varies by country (more information on exit in SI part 4).
}
\section{Discussion}


{Legal uncertainty and prolonged waiting times undermine the psychosocial and physical health of refugees, impeding their integration into society \cite{nagel2023stuck, phillimore2021violence}. Despite this, the legal journeys of migrants and refugees are often overlooked in quantitative migration studies, with much of the existing research focusing on success rates and rejections rather than the complexities of legal transitions~\cite{eule2018migrants,storato2023timescapes}. While basic statistics on outcomes such as asylum acceptance or rejection are available, the specific pathways migrants follow from entry to stable legal status remain largely unexplored. This study examines the legal status transitions in Austria over two years. By modeling how migrants move between legal statuses, we can predict future behaviors and observe how legal journeys differ across countries. Our findings reveal that migrants from different countries follow distinct trajectories, influenced by nationality, conflict background, and legal pathways. These distinct legal trajectories can be seen as signatures that characterize the migration and legal experiences unique to each group. These insights enhance our understanding of migration processes and provide valuable information for refining migration policies.
}

{
Our results reveal that legal trajectories differ systematically across countries of origin, reflecting the broader structural and legal frameworks within which migrants and refugees are processed. Migrants from EU member states, such as Germany, Italy, and Croatia, follow comparatively simple trajectories, moving through few legal statuses and reaching stable residence quickly (more information in SI part 3). This is consistent with the rights conferred by the EU Directive 2004/38/EC, under which EU nationals are not required to seek authorisation to reside and face substantially lower administrative barriers~\cite{Directive2004_38}. By contrast, migrants from conflict-affected regions such as Syria and Afghanistan experience fragmented and prolonged trajectories with significantly more transitions. For these populations, the path to stability runs through the asylum system, where refugee status under the Geneva Convention requires individual proof of persecution, and subsidiary protection applies to those facing serious harm but not meeting that individual threshold~\cite{geneva-refugee-convention,LukasAIDA2024}. The prevalence of ``entitled to asylum'' and eligible for ``subsidiary protection'' statuses among Syrian and Afghan migrants reflects both the nature of their displacement and the demands of these evidentiary processes, compounded by administrative backlogs and resource constraints. Migrants from Ukraine follow a distinct trajectory, predominantly reaching ``displaced person'' status through a comparatively shorter process. This is directly attributable to the activation of the EU Temporary Protection Directive in March 2022, which granted Ukrainians immediate and collective protection without requiring individual asylum applications~\cite{TempAIDA2024}.
}

{
The study also reveals important insights into the time it takes for migrants to reach a stable legal status. As expected, migrants from countries such as Germany and Italy experience the shortest wait times, with Germans obtaining stable residence status in under four months. In contrast, migrants from countries like Afghanistan face the longest wait times, averaging more than twenty months. This difference in timelines highlights the administrative challenges refugees from conflict zones face and the potential strain it places on their integration process. Reducing wait times for refugees, particularly those from countries like Afghanistan and Syria, could help improve their long-term integration prospects.
}

{The mode of entry, whether individuals arrive directly as asylum seekers or first register as foreigners, shapes legal outcomes in ways that extend beyond the asylum determination itself. Our results show that those who enter directly as asylum seekers face higher exit rates and lower chances of obtaining stable protection status within the first year, compared to those who first enter as foreigners before initiating an asylum claim. While this gap may partly reflect compositional differences between the two groups, for instance differences in country of origin or individual circumstances, it also points to a structural effect of entry status on subsequent legal trajectories, raising questions about consistency and equity across pathways into the Austrian legal system.}

{
For some countries, we find a slight female advantage in the legal journey, which aligns with recent findings in both asylum application outcomes, where women show a slight advantage \cite{ortensi2024female}, and housing, where male refugees experience greater residential instability \cite{stability_Refugee}. This may be linked to who initiates the asylum process and bears the burden of establishing a presence in the host country. In contexts like Syria and Afghanistan, where men are often the first to flee and apply for asylum \cite{demarchi2019gender}, rejection rates tend to be higher. Women may arrive later through family reunification programs, increasing their chances of acceptance. Conversely, in Ukraine, the initial refugee population is predominantly female, which may lead to higher acceptance rates for the men who later join them as part of family reunification.
}
{Our findings reveal persistent disparities in legal status trajectories across migrant groups. These disparities suggest that existing procedures may inadvertently privilege some groups over others, even when their legal needs and vulnerabilities are comparable. For example, migrants from Ukraine rapidly attain ``displaced person'' status through a streamlined process, while those from Afghanistan or Syria face lengthier and more uncertain pathways despite similarly urgent protection needs. Possible reasons for the observed delays and complexities in legal transitions may include the pressure on caseworkers handling asylum cases to prioritize the quantity of resolved asylum applications over their quality, which can result in reduced interaction with asylum seekers and hinder the accuracy of legal assessments. Moreover, a lack of specialized legal education may limit their ability to handle complex cases effectively \cite{Dahlvik2018}. Our findings highlight the need for clearer administrative procedures to reduce disparities in legal transitions. Standardizing timelines, clarifying eligibility for legal statuses, and improving legal counsel access during application processes would help reduce uncertainty and bias \cite{schittenhelm2019implementing, eagly2016access}. 
}

{While the model predicts legal transitions, it cannot account for sudden policy changes, as the transition probabilities can only be updated retrospectively. In addition, we do not incorporate factors such as family reunification, educational attainment, and employment, which may shape legal outcomes. Although residential postal codes could proxy for the relevant local authority, they do not guarantee precision in identifying the processing branch. Whether variation across branch offices or regional directorates accounts for some of the observed differences, therefore, remains an open question \cite{bfa2026, aida2025admissibility}. The dataset captures registered legal status but not the intent or circumstances behind individuals' legal trajectories; the reasons why individuals follow particular pathways, whether structural, strategic, or circumstantial, remain unobservable from administrative records alone. Our two-year observation window limits the ability to capture longer-term patterns, including repeated reentry, complete trajectories, and possible cyclical procedures that refugees and asylum seekers may experience~\cite{eule2018migrants}. Future research should address these limitations by refining transition models to incorporate individual-level covariates, longer observation periods, and spatial heterogeneity in institutional processing.
}

{Our analysis is anchored in Austria due to the availability of high-resolution, daily administrative data and its central role in hosting refugee populations. The legal trajectories we observe, rapid stabilization for some groups and prolonged uncertainty for others, reflect dynamics that are not unique to Austria. Comparable patterns appear in other EU countries such as Germany and Italy \cite{fontanari2017s,kosyakova2020role}. In contrast, the United States follows a quota-based system in which refugees are supported by resettlement partners rather than directly by the state \cite{UNHCR_US_resettlement_partners}. Despite these institutional differences, the central challenge of achieving timely legal stability for refugees remains unresolved.
}

{This study demonstrates the potential to improve migration governance by offering a dynamic, system-wide view of legal status transitions. Tracking changes in legal status across time and populations allows identifying bottlenecks in the system, such as extended stays in transitionary statuses or high rates of legal exits following rejections. On the individual level, monitoring deviations from expected transition trajectories can support caseworkers in quickly identifying individuals whose legal status has stalled beyond expected durations, enabling timely intervention before delays compound. The 2022–2024 transition probabilities establish a pre-intervention baseline against which the effects of the EU Asylum and Migration Pact, applicable from June 2026, can be directly measured once implemented~\cite{oesterreich2026asylum}.
Additionally, regularly analyzing such data could allow for early warning systems that flag patterns of unequal treatment. By embedding data-driven modeling, governments can move beyond reactive responses and toward proactive, equitable management of migration and asylum systems.

}
\section{Methods}

\subsection{Data and the Legal System }

{This study uses individual-level administrative records provided by the Federal Ministry of  the Interior of Austria (\textit{Bundesministerium f\"{u}r Inneres}, BMI), which maintains and safeguards the \textit{Zentrales Melderegister} (ZMR --- Central Register of Residents). The ZMR records the legal residence status of every registered non-Austrian national on a 
daily basis and constitutes one of the most granular administrative sources available for studying legal trajectories in Austria. The same administrative records have previously been used to study international migration flows~\cite{prieto2023diaspora} and refugees' internal mobility~\cite{stability_Refugee}. The present study extends 
this line of work by analyzing the temporal legal status transitions, rather than the spatial movement of migrants and refugees. The data are not publicly available. The BMI safeguards the dataset and made it accessible to our research institution under a formal research agreement and strict data protection regulations. 

In Austria, asylum cases are processed through the Federal Office for Immigration and Asylum (BFA), which operates via regional offices in each of the nine federal provinces and their branch offices~\cite{bfa2026,aida2025admissibility}. Residence permit applications are handled by separate settlement authorities at the provincial or municipal level, for example, in Vienna, by Municipal Department 35~\cite{oesterreichgvat_renewal_residence_permits_2026}. Our dataset records legal statuses over time but does not identify the specific office, branch, or caseworker responsible for each case. Other variables (for example, residential postal codes) reflect registered place of residence rather than point of entry and thus do not guarantee precision in identifying the processing branch.
}
{
Third-country nationals who plan to reside or settle for more than six months are required to obtain a residence permit, specific to their purpose of stay (e.g., family reunification, study, or employment)~\cite{BMEIA2024}. Individuals from the EU are not subject to this requirement and may remain as foreigners without needing a residence permit. For asylum seekers, obtaining a visa in their country of origin can be challenging. As a result, some asylum seekers enter Austria irregularly, without a visa, and apply for asylum upon arrival~\cite{iom2023,costello2016eu}.
}

{
The system classifying foreigners in Austria has 11 legal statuses (Table~\ref{tab:legal_classes}). We introduce an additional status, ``exit'', to account for individuals who have legally exited Austria, bringing the total to 12 legal statuses. Status changes are assessed for each individual and recorded daily at the country level. If any status changes occur within a given day, they are documented, providing a daily summary of how the legal status of individuals from each country has changed. We group the legal statuses into four main categories, with each documented migrant falling into one of these statuses. Entry-level status applies to newly arrived migrants who are classified as foreigners. Transitionary statuses refer to individuals in the process of obtaining a stable legal status or awaiting departure, such as those who are seeking asylum or tolerated foreigners. Stable and long-term statuses refer to individuals who have obtained the right to live and work in Austria, including those with residence permits or asylum entitlement. Finally, exit-related statuses include individuals who have either left or are in the process of leaving Austria, such as those voluntarily departing, being obligated to leave, or facing deportation.
\begin{table}[!ht]
\centering
\begin{tabular}{|p{4cm}|l|>{\centering}m{0.8cm}|c|p{4cm}|}
 \hline
 \textbf{Migrant Legal Status} & \textbf{Classification} & \textbf{Icon} & \textbf{Frequency (\%)} & \textbf{Description} \\
 \hline
Foreigner & Entry & \includegraphics[width=0.6cm]{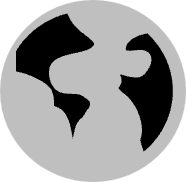} & 19.34 & Any non-Austrian national subject to entry and stay regulations under NAG* \\
\hline
Settled migrant with residence permit & Stable & \includegraphics[width=0.6cm]{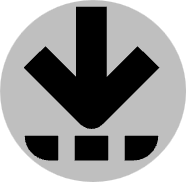} & 42.86 & Holder of documents granting long-term residence under NAG* \\
Entitled to asylum & Stable & \includegraphics[width=0.6cm]{Figures/icons/stable.png} & 5.04 & Granted full asylum status after positive decision in the asylum procedure, under AG** \\
Displaced person & Stable & \includegraphics[width=0.6cm]{Figures/icons/stable.png} & 6.59 & Persons who have been forced to flee their places of habitual residence, granted protection under AG** \\
Eligible for subsidiary protection & Stable & \includegraphics[width=0.6cm]{Figures/icons/stable.png} & 0.95 & Granted protection when serious harm is faced on return, under AG** \\
Humanitarian residence permit & Stable & \includegraphics[width=0.6cm]{Figures/icons/stable.png} & 0.05 & Discretionary permit to stay on humanitarian grounds under NAG* \\
\hline
Approved asylum seeker & Transitionary & \includegraphics[width=0.6cm]{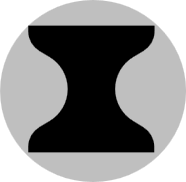} & 4.81 & Asylum seeker whose application was accepted and is waiting final decision under AG** \\
Asylum seeker & Transitionary & \includegraphics[width=0.6cm]{Figures/icons/transition.png} & 0.23 & Foreign national who placed an asylum application waiting acceptance under AG** \\
Tolerated foreigner & Transitionary & \includegraphics[width=0.6cm]{Figures/icons/transition.png} & 0.26 & Stay formally tolerated when deportation is impossible, under FPG*** \\
\hline
Foreigner with obligation to leave the country & Exit & \includegraphics[width=0.6cm]{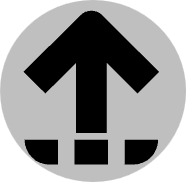} & 0.25 & Subject to a formal departure order under FPG*** \\
Manually inhibited foreigner with obligation to leave the country & Exit & \includegraphics[width=0.6cm]{Figures/icons/exit.png} & 0.01 & Deportation is suspended due to an entry prohibition until the ban expires \\
Exit & Exit & \includegraphics[width=0.6cm]{Figures/icons/exit.png} & 19.60 & Recorded departure via deregistration of residence or death \\
 \hline
\end{tabular}
\caption{Distribution of migrant legal residence statuses, their classification, frequency, and brief description (total observations = 140,372,370). \\
*Niederlassungs- und Aufenthaltsgesetz (NAG): law regulating the residence of foreigners in Austria who stay for a longer period of time. \\
**Asylgesetz (AG): law defining asylum seekers, asylum status, and subsidiary protection.\\
***Fremdenpolizeigesetz 2005 (FPG): law covering toleration, departure orders, entry bans. }
\label{tab:legal_classes}
\end{table}
}

{
In our analysis, we focus on new arrivals who entered from November 2022 onward, as these are the majority of migrants experiencing a legal status change. We take into consideration migrants from the top 20 countries of origin, which is more than 85\% of the total migrant population, and we observe them for two years. During this observation period, more than 350,000 migrants and refugees entered Austria, with a total of 140,372,370 legal status entries, with varying frequencies for each category (Table~\ref{tab:legal_classes}).

Exit is considered when a person permanently leaves the country. We do not allow for re-entry in our analysis. If a migrant leaves and returns within the two years of observation, we treat this as a continuation of their original stay. Naturalization and death, although unlikely (during 2022--2024 naturalization rate remained below 1\% in Austria \cite{ statistikaustria}), are considered in the exit state. Therefore, exit is a permanent or absorbing status in our analysis.   

We observe migrants who are forced to leave their country to seek refuge and apply for asylum status; we refer to them as \textit{Refugees}. These are individuals who undergo the process of seeking asylum, regardless of whether they are successful or not.  
We find that remaining in the same status has the highest probability, meaning that on a daily basis, changing legal status could be considered a rare event. Yet, collectively, over our 734-day observation period covering 358,327 people, we observe 367,146 legal switches daily with an average of 1.02 switches per person (more information in SI part 1). 
}

{
The majority of individuals engaged in the legal process are of working age. However, regions affected by conflict show a higher proportion of minors among migrants (more information in SI part 2). For example, nearly 50\% of newcomers from Syria are minors, illustrating the demographic impact of conflict on migration patterns.
}

{
To estimate how many migrants changed their legal status, we identify individuals who experienced at least one status change during the observation period. We use a daily description of the data as it is the most granular observation unit available. 

The choice of the time interval is relevant to our analysis. Using daily intervals, we find that around 70\% of the population changes their legal status at least once (Table \ref{tab:stayers_changers}). In contrast, aggregating the data to weekly intervals results in the loss of some observations and reduces the proportion of those who change their legal status at least once to around 30\%. Therefore, we choose to proceed with daily intervals. 
{
\begin{table}[!ht]
\centering
\begin{tabular}{|l|r|r|r|r|}
\hline
\textbf{Country} & \textbf{Stayer} & \textbf{Changer} & \textbf{Total} & \textbf{Changer \%} \\
\hline
Germany        & 14,527 & 29,943 & 44,470 & 67.3 \\
Romania       & 14,965 & 25,107 & 40,072 & 62.7 \\
Ukraine        & 5,290  & 33,378 & 38,668 & 86.3 \\
Hungary       & 12,722 & 24,292 & 37,014 & 65.6 \\
Syria          & 6,395  & 26,830 & 33,225 & 80.8 \\
Turkey          & 6,726  & 13,088 & 19,814 & 66.1 \\
Croatia        & 6,251  & 13,200 & 19,451 & 67.9 \\
Serbia          & 6,078  & 10,938 & 17,016 & 64.3 \\
Bosnia and Herzegovina  & 4,675 & 10,604 & 15,279 & 69.4 \\
Slovakia        & 4,789  & 8,579  & 13,368 & 64.2 \\
Italy          & 3,617  & 7,961  & 11,578 & 68.8 \\
Poland          & 4,140  & 6,447  & 10,587 & 60.9 \\
Bulgaria        & 3,397  & 6,569  & 9,966  & 65.9 \\
Afghanistan     & 2,340  & 6,769  & 9,109  & 74.3 \\
Russia       & 2,001  & 6,292  & 8,293  & 75.9 \\
China         & 2,080  & 4,541  & 6,621  & 68.6 \\
India          & 1,669  & 4,929  & 6,598  & 74.7 \\
Iran          & 1,494  & 4,869  & 6,363  & 76.5 \\
Kosovo          & 1,907  & 4,003  & 5,910  & 67.7 \\
North Macedonia  & 1,450 & 3,475  & 4,925  & 70.6 \\
\hline
\textbf{All} &106,513 & 251,814& 358,327&\textbf{70.27} \\
\hline
\end{tabular}
\caption{Distribution of migrants who do not change their legal status during the observation period (Stayers) and those who do (Changers) across countries, along with the total number of new entries and the percentage of migrants who change their legal status (Changer\%).}
\label{tab:stayers_changers}
\end{table}
}
}


\subsection{Legal and Institutional Context}


{
The legal framework governing residence in Austria operates under fundamentally different logics for EU and non-EU nationals. EU citizens (including EEA nationals and Swiss citizens) derive their right to reside directly from EU Free Movement Directive — specifically Directive 2004/38/EC, Article 7~\cite{Directive2004_38}. In practice, EU nationals may remain in Austria for up to three months without any conditions; residence beyond three months is permitted provided they are employed, self-employed, enrolled in education, or hold sufficient financial resources (under the Aliens Police Act (FPG), and under §9, §51, §53 and §53a of Settlement and Residence Act (NAG))~\cite{EU_2026}. The same framework extends to qualifying family members of EU citizens who are themselves EU nationals~\cite{EUFamily_2026}. Stays exceeding three months must be reported to the settlement authority within four months of entry, after which a registration certificate will be issued upon request. This document confirms a pre-existing right rather than constituting authorization to reside. 

Third-country nationals should generally apply for and receive authorization to reside before or upon entry, with permits issued for the purpose of their stay (e.g., family reunification, study, employment) for initial application~\cite{NonEU_first_2026}. Applications are assessed on an individual basis and are contingent on the applicant's ability to demonstrate sufficient income, health insurance, and adequate housing (under §11 of the Settlement and Residence Act (NAG))~\cite{NonEU_2026}. Unlike EU nationals, whose right to reside arises by operation of law, third-country nationals must obtain explicit authorization, generating administrative processing times that have no structural equivalent for EU nationals.

This structural bifurcation means that observed differences in transition speed and trajectory complexity between EU and non-EU populations are, in part, a direct product of the underlying legal architecture. Our analysis treats EU and non-EU origin groups separately where appropriate (particularly in SI part 3), while acknowledging that the ``non-EU'' category obscures substantial variation across labor migrants, family migrants, and asylum seekers subject to distinct procedural tracks.

}

{
For third-country nationals seeking asylum, there are four protection statuses that constitute stable outcomes:  entitled to asylum, displaced person, eligible for subsidiary protection, and humanitarian residence permit (Table~\ref{tab:legal_classes}). The four statuses differ substantially in the rights they confer and in access to family reunification. Entitled to asylum, granted under the Geneva Convention, provides the broadest protection package: unlimited residence, full and immediate labor market access, and an immediate right to family reunification (under §3 of the Asylum Act  (AsylG))~\cite{geneva-refugee-convention,LukasAIDA2024,General_Asylum2026}. 

Eligible for subsidiary protection, is granted for individuals who do not qualify for refugee status under the Geneva convention, but would face a threat to their life or physical integrity  upon return to their country (under §8 AsylG)~\cite{General_Asylum2026}. Their residence is limited and subject to renewal and they have full access to labor market. Family reunification is additionally subject to a three-year waiting period (§35 AsylG) \cite{General_Asylum2026,LukasAIDA2024}.

Displaced persons, a status activated via the Temporary Protection Directive by implementing decision (EU) 2022/382 and applied primarily to Ukrainians from 2022, is limited in duration and granted collectively rather than through individual determination. Unlike the other two statuses, it confers freedom of movement within Austria and onward to other EU member states, though extended absences risk status expiry. It guarantees access to the labour market as well as healthcare and basic social support. However family reunification is restricted to the narrow framework of the Temporary Protection Directive. The status currently runs until at least 4 March 2027~\cite{Displaced_2026,TempAIDA2024}

The humanitarian residence permit is a non-EU-harmonized national protection status (§§56--57 AsylG), unlike the first three~\cite{bassermann2019}. It has two distinct variants: §56 is integration-oriented, granted to long-term residents meeting presence and integration conditions; §57 is protection-oriented, covering tolerated stays, crime victims, and domestic violence cases. Both variants are valid for 12 months and non-renewable as such, labour market access is conditional on the specific variant held, and neither entitles the holder to family reunification or guaranteed social assistance.
}

\subsection{Constructing The Transitions Between Legal Status}

{
In our dataset, we observe direct switches from one legal status to the next. However, since we only have two years of observation, we do not observe many complete trajectories (until a person has exited the country). By assuming the Markovian property, where each transition depends only on the current state, we can reconstruct the likely pathways migrants follow across statuses, effectively inferring the full legal journey even with partial data.

Markov processes have been used in social science to study the progression of individuals through different states, either migration states or criminal states \cite{Wolfganag_Crimeinbirth,curiel2022mobility}. A Markov Process is a sequence of random variables where the conditional probability of the future state depends only on the present state, not on the sequence of events that preceded it \cite{parzen1999stochastic}:
\begin{align*}
    P(X_{t+1}|X_t=x_t,X_{t-1}=x_{t-1},\dots, X_1= x_1 =  \\
    P(X_{t+1}=x_{t+1}|X_t = x_t),
\end{align*}
where $X_t$ represents the state at time $t$ and $x_t$ is a possible state in the set of states.

In our study, we compile daily transitions for each individual where each individual's state at day~$t$ is denoted as $X_t$ and belongs to a set of $N=12$ possible legal statuses. For a given migrant, the transition from one state to another is observed for two years, where the different legal states such that $X_t \in{\{1,\dots 12\}}$ represent the status of a migrant at day $t$.  

For each group, representing migrants from a certain country of origin or gender, the transition matrix $T$ is a $12 \times 12$ matrix constructed for each pair of states $i$ and $j$, 
\begin{equation*}
    T_{ij} = P(X_{t+1} = j | X_t = i) = \frac{C_{ij}}{\sum_{k} C_{ik}},
\end{equation*}
where $C_{ij}$ is the count of transitions from state {i} to state {j}.
}

\subsection{Policy Context and Limitations}
{
The transition matrices estimated here assume stationary transition probabilities across the observation window, a constraint that becomes consequential in light of a specific mid-period policy shift. In May 2024, the Ministry of the Interior interrupted ongoing family reunification cases and introduced stricter verification requirements, including more systematic use of DNA testing and re-examination of previously approved applications; processing effectively halted between May and August/September 2024, with monthly arrivals via this channel falling sharply~\cite{LukasAIDA2024}. In 2025, the Austrian parliament further enacted legislation authorizing the government to suspend new family reunification applications by refugees altogether~\cite{LukasAIDA2024, parlamentoesterreich2025pause}.

Family reunification is one of the pathways through which individuals transition into settled residence status in our dataset. The May 2024 disruption temporarily suppressed this transition for individuals whose route to stability ran through this channel, primarily affecting Afghan and Syrian nationals. Because this change occurs at month 18 of the 24-month observation window, its weight in the overall estimated transition matrix is modest relative to the preceding 18 months. It may nonetheless marginally bias downward the estimated probability of reaching stability for affected groups in the final period, and the post-observation legislative suspension means the transition probabilities reported here reflect a procedural regime that no longer holds~\cite{UNHCR_Austria_FamilyReunification_2025}. Future work using time-varying transition matrices would be better placed to disentangle the effects of such mid-period shifts on observed trajectories.
}
\clearpage
\printbibliography

\section*{Acknowledgements}
We thank Markus Hofer for his insightful input on the theoretical aspects of the model and Dr. Ljubica Nedelkoska, as well as the visiting members of the Society Moving lab at the Complexity Science Hub, for their valuable comments. Special thanks to Liuhuaying Yang and the visualization team at the Complexity Science Hub for their contributions.


\section*{Competing interests}
The authors declare that they have no competing interests.

\section*{Author's contributions}
\begin{itemize}
    \item Conceptualization: O.A., R.P.C
    \item Data curation: R.S.
    \item Formal Analysis: O.A.
    \item Methodology:  O.A., R.P.C
    \item Visualization: Figure 1-4  O.A.,  Figure 5 E.D. 
    \item Writing - original draft: O.A. 
    \item Writing - review \& editing: O.A., E.D., G.P.V., R.S., A.V., C.K., R.P.C. 
\end{itemize}
\section*{Funding}
This research is funded by the Federal Ministry of the Interior of Austria (2022-0.392.231).

\section*{Data availability}
The raw and processed data are not available due to privacy laws. The Federal Ministry of the Interior of Austria safeguarded the dataset and made it accessible to our research institution under strict data protection regulations. Researchers must reach individual agreements with the Ministry to access the data.

\pagebreak

\clearpage
\section{Supplementary Information}

\section{Markov Assumption and Transition Frequency}

{
Our analysis is based on the Markov assumption, which states that the legal status of a person depends only on the previous status. Testing the validity of the Markov assumption requires historical data on the legal switches covering, on average, many switches~\cite{stander1989markov}. However, our dataset does not support this, given that the average number of status changes per person is approximately one (Table~\ref{tab:country_switches}), with no group exceeding 1.5 transitions on average. Even among migrants from Iran, Russia, and Afghanistan, who exhibit slightly higher rates, the data remain insufficient for empirical validation. This limitation makes it infeasible to formally test the Markov assumption.

\begin{table}[!ht]
\centering
\begin{tabular}{|l|r|}
\hline
\textbf{Country} &  \textbf{Mean Switch Count} \\
\hline
               Germany &               0.95 \\
               Romania &               0.88 \\
               Ukraine &               1.22 \\
               Hungary &               0.91 \\
                 Syria &               1.17 \\
                Turkey &               1.10 \\
               Croatia &               0.91 \\
                Serbia &               0.97 \\
Bosnia and Herzegovina &               0.98 \\
              Slovakia &               0.91 \\
                 Italy &               1.02 \\
                Poland &               0.86 \\
              Bulgaria &               0.92 \\
           Afghanistan &               1.25 \\
                Russia &               1.32 \\
                 China &               0.99 \\
                 India &               1.16 \\
                  Iran &               1.33 \\
                Kosovo &               1.09 \\
       North Macedonia &               1.06 \\
     \hline
     \textbf{All Countries } & \textbf{ 1.02} \\
\hline
\end{tabular}
\caption{The average number of legal switches observed in the dataset by country of origin.}
\label{tab:country_switches}
\end{table}
}
\

\section{Gender and the legal transitions}
\label{SI:Gender}

{
In the data, we have almost the same number of females and males (Table~\ref{tab:Gender}) except for countries with ongoing conflict like Ukraine, Syria, and Afghanistan. In our analysis, when considering gender alone, we observe only minor differences in the probability of settling after one year (Fig.~\ref{fig:Gender}). The likelihood of obtaining a stable residence is nearly identical, with 50.2\% for men and 50.7\% for women. However, a small gender gap appears when considering the probability of gaining any refugee status: this is around 15\% for women and 11\% for men. Men also show a marginally higher tendency to leave after one year, with a 26.6\% probability compared to 25.42\% for women. On average, men take slightly longer to reach stability: 5.4 months versus 4.6 months for women.

These differences become more pronounced when we examine the intersection of gender and country of origin. In general, women have a higher probability of gaining residence and a slightly lower chance of leaving. While these differences are relatively small in countries such as Germany (residence probability: 75\% for men vs. 79\% for women), Turkey (48\% for men vs. 53\% for women), and Serbia (37\% for men vs. 43\% for women). It becomes more pronounced in conflict-affected countries like Syria (2.4\% for men vs. 13\% for women), and Afghanistan (14\% for men vs. 50\% for women). In contrast, for Ukraine, the probability of gaining residence is negligible and almost identical for both genders (1.46\% for men vs. 1.47\% for women).

The probability of gaining asylum is consistently higher for women across our analysis. For instance, among Syrian refugees, women have a 65\% probability of obtaining asylum compared to 46\% for men (please check the interactive version for the country gender specific plots here: \url{https://vis.csh.ac.at/migration-legal-status/}). For Afghani refugees, women have a 20\% probability of obtaining asylum versus 14\% for men. Notably, Afghan women also reach asylum faster, with an average of 14 months compared to 30 months for men. In contrast, we observe the opposite trend for people coming from Ukraine; the probability of obtaining displaced person status is slightly lower for women (67\%) compared to men (73\%). 

These disparities may be linked to who initiates the asylum process and bears the burden of relocating to the host country, rather than the gender of the individuals. For countries like Syria and Afghanistan, where men are often the first to flee and apply for asylum~\cite{demarchi2019gender}, rejection rates tend to be higher. Women may arrive later through family reunification programs, increasing their chances of acceptance. Conversely, in Ukraine, the initial refugee population is predominantly female, which may lead to higher acceptance rates for the men who later join them as part of family reunification. 

\begin{table}[ht!]
\centering
\begin{tabular}{|l|r|r|}
\hline
\textbf{Country} & \textbf{Female \%} & \textbf{Male \%} \\
\hline
Germany & 50.2 & 49.8 \\
Romania & 47.1 & 52.9 \\
Ukraine & 57.9 & 42.1 \\
Hungary & 46.9 & 53.1 \\
Syria & 45.8 & 54.2 \\
Turkey & 40.7 & 59.3 \\
Croatia & 46.6 & 53.4 \\
Serbia & 45.9 & 54.1 \\
Bosnia and Herzegovina & 48.6 & 51.4 \\
Slovakia & 49.5 & 50.5 \\
Italy & 46.9 & 53.1 \\
Poland & 41.0 & 59.0 \\
Bulgaria & 49.1 & 50.9 \\
Afghanistan & 35.9 & 64.1 \\
Russia & 53.8 & 46.2 \\
China & 46.4 & 53.6 \\
India & 39.0 & 61.0 \\
Iran & 50.9 & 49.1 \\
Kosovo & 42.9 & 57.1 \\
North Macedonia & 46.7 & 53.3 \\
\hline
\textbf{All} &\textbf{47.8}&\textbf{52.2} \\
\hline
\end{tabular}
\caption{Gender compositions of migrants who entered Austria during our observational period.}
\label{tab:Gender}
\end{table}

\begin{figure}[!ht]
  \centering
  \includegraphics[width=\textwidth]{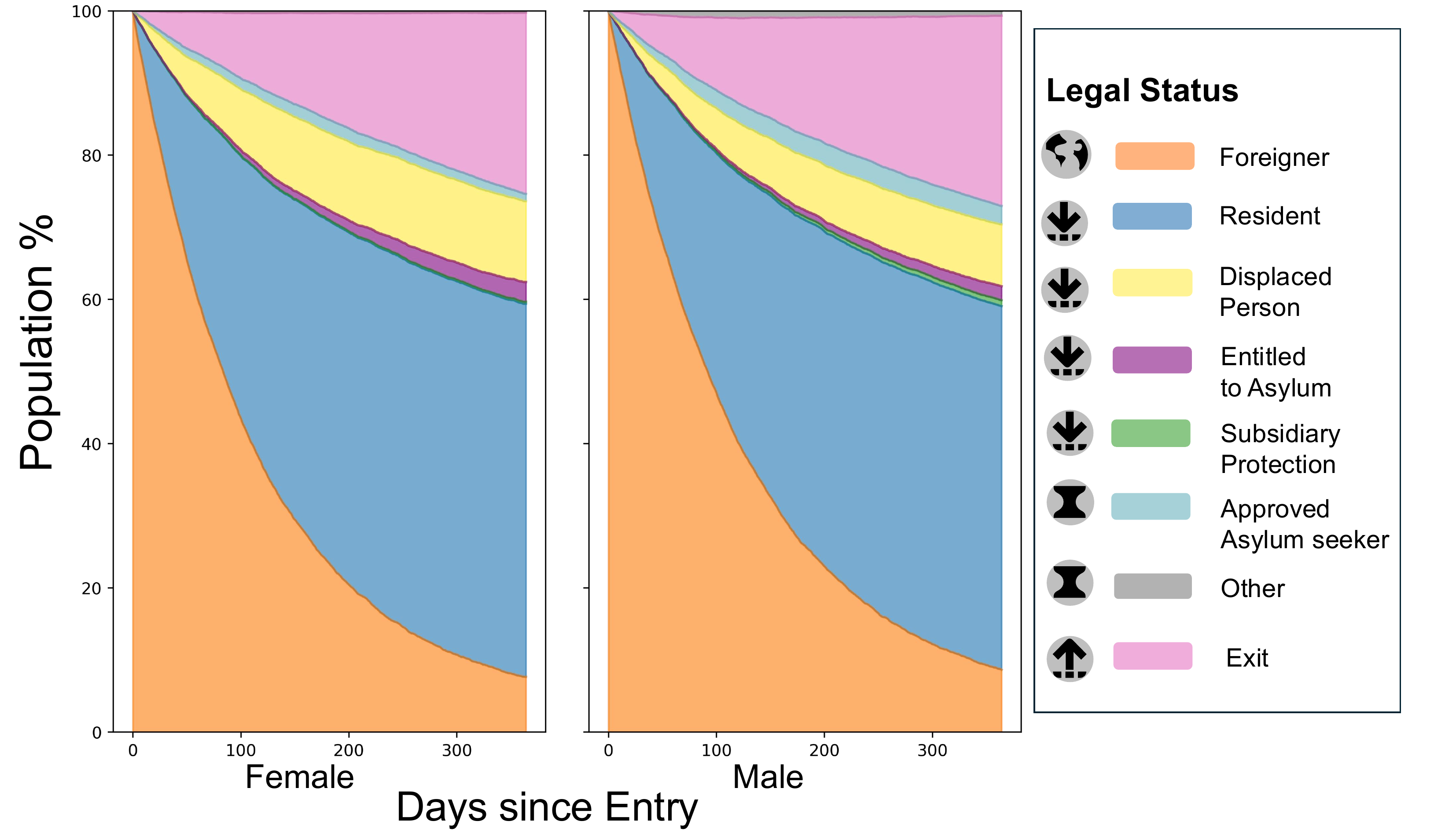}
  \caption{\textbf{Gender Differences.} Modeled share of 10,000 female and male migrants transition through different legal statuses. The probability of attaining stability within the first year is estimated based on observed transition patterns. An interactive version of this figure and all country and gender specific plots are available at:  \url{https://vis.csh.ac.at/migration-legal-status/}.}
  \label{fig:Gender}
\end{figure}

\clearpage
{
The age composition of different migrant groups (Table~\ref{tab:country_AgeGroup}) provides additional context for interpreting these gendered patterns, as countries with a higher share of minors or older adults may have different vulnerabilities and family structures that influence who migrates first, who applies for asylum, and how legal stability is eventually reached.
\begin{table}[ht!]
\centering
\begin{tabular}{|l|r|r|r|}
\hline
\textbf{Country} &  \textbf{0--17 \%} &  \textbf{18--64 \%} &  \textbf{65+ \%} \\
\hline
Germany & 13.6 & 81.3 & 5.1 \\
Romania & 24.9 & 73.1 & 1.9 \\
Ukraine & 22.9 & 69.5 & 7.6 \\
Hungary & 18.1 & 80.3 & 1.5 \\
Syria & 54.0 & 45.7 & 0.3 \\
Turkey & 25.6 & 69.5 & 4.9 \\
Croatia & 27.5 & 71.0 & 1.5 \\
Serbia & 24.3 & 71.1 & 4.6 \\
Bosnia and Herzegovina & 21.7 & 75.4 & 2.9 \\
Slovakia & 19.4 & 78.3 & 2.3 \\
Italy & 13.7 & 84.5 & 1.7 \\
Poland & 18.6 & 79.1 & 2.3 \\
Bulgaria & 23.9 & 73.1 & 2.9 \\
Afghanistan & 34.7 & 64.9 & 0.5 \\
Russia & 22.3 & 73.2 & 4.5 \\
China & 11.6 & 83.7 & 4.7 \\
India & 14.0 & 81.0 & 5.1 \\
Iran & 15.0 & 78.8 & 6.2 \\
Kosovo & 28.8 & 67.6 & 3.6 \\
North Macedonia & 32.4 & 64.2 & 3.5 \\
     \hline
     \textbf{Total} & \textbf{24.26} & \textbf{72.39} & \textbf{3.35}\\
\hline
\end{tabular}
\caption{The age composition of each country of origin, the majority of the countries are in the working age group, except for Syria, where close to half are minors.}
\label{tab:country_AgeGroup}
\end{table}
}

}
\section{EU Membership}
\label{SI:EU}

{
Although grouping migrants into broad categories such as EU vs. non-EU can highlight general patterns, we do not rely on them in our analysis, as they obscure substantial variation at the country level. For example, non-EU countries include Serbia, China, Russia, Syria, Afghanistan, and Ukraine. Grouping them does not reflect meaningful similarities in legal trajectories or outcomes relevant to our study.

EU migrants follow a more straightforward legal trajectory, typically entering as foreigners, obtaining residence permits, and eventually leaving the country, a pattern consistent with the free movement rights established under Directive 2004/38/EC~\cite{Directive2004_38} (for a full discussion, see Section 4.2). Within the first year, the probability of obtaining a residence permit is 68\%, while the probability of leaving is 31\%. An interactive visualization of EU probabilities is available at: \url{https://vis.csh.ac.at/migration-legal-status/}.
}

\section{Time to Exit}

{
Time to exit the country varies for different countries of origin (Table~\ref{tab:Exit_Time}). To estimate it, we calculate the mean first passage time from the entry state ``foreigner'' to the exit state, also known as the hitting time. Given the probability matrix $T$, we solve the linear system $(I-T+\varepsilon I) h = b$. Where $h$ is the hitting time and is equivalent to the number of steps needed to reach that state, $ \varepsilon$ is a small constant we added to ensure numerical stability, $\varepsilon = 10^{-6}$, $b$ is a vector with all entries set to $1$, except for the index corresponding to the exit state, which is set to $0$, and $I$ is the identity matrix. 
\begin{table}[ht!]
    \centering
    \begin{tabular}{|l|r|}
        \hline
        \textbf{Country} & \textbf{Mean First Passage Time (in years) }\\
        \hline
        Germany & 6.4 \\
        Romania & 4.0 \\
        Ukraine & 3.4 \\
        Hungary & 4.3 \\
        Syria & 30.9 \\ 
        Turkey & 3.1 \\ 
        Croatia & 7.3 \\
        Serbia & 3.2 \\ 
        Bosnia and Herzegovina & 3.9 \\
        Slovakia & 3.3 \\
        Italy & 4.1 \\
        Poland & 5.0 \\
        Bulgaria & 4.1 \\
        Afghanistan & 8.7 \\
        Russia & 6.1 \\
        China & 2.4 \\
        India & 4.6 \\
        Iran & 3.8 \\ 
        Kosovo & 4.6 \\
        North Macedonia & 5.3 \\
        \hline
    \end{tabular}
    \caption{Mean first passage time (in years) from the state ``foreigner'' to exiting the country.}
    \label{tab:Exit_Time}
\end{table}
}

\end{document}